\documentclass[sigconf]{acmart}

\usepackage{amsmath}
\usepackage{mathtools}

\usepackage{xcolor}
\usepackage{tikz}
\usetikzlibrary{arrows.meta,positioning,calc}

\acmYear{2026}
\acmConference[DAI '26]
  {The Eighth International Conference on Distributed Artificial Intelligence}
  {November 29--December 2, 2026}
  {Hong Kong, China}
\acmBooktitle{The Eighth International Conference on Distributed Artificial
  Intelligence (DAI '26), November 29--December 2, 2026, Hong Kong, China}

\setcopyright{none}
\copyrightyear{}
\acmDOI{}
\acmISBN{}
\newtheorem{assumption}{Assumption}
\newcommand{\E}{\mathbb{E}}
\newcommand{\Prob}{\mathbb{P}}
\newcommand{\CANO}{\textsc{Cano}}
\newcommand{\ind}{\mathbf{1}}

\newcommand{\ReferenceMStar}{13}
\newcommand{\ReferenceQStar}{0.161}
\newcommand{\ReferenceCapStar}{0.648}
\newcommand{\ReferenceProfit}{0.465}
\newcommand{\GridScenarioCount}{1944}

\newcommand{\EstimationLargeRegret}{0.0033}

\newcommand{\MonteCarloError}{0.00040}

\title[Parallelism or Concession?]{Parallelism or Concession? Concurrency-Aware Procurement Negotiation for Agentic Commerce}

\author{Xiaolin Xu}
\orcid{0000-0003-0598-657X}
\affiliation{%
  \department{School of Business}
  \institution{Nanjing University}
  \streetaddress{16 Jin Yin Street, Gulou District}
  \city{Nanjing}
  \state{Jiangsu}
  \postcode{210093}
  \country{China}}
\email{xuxl@nju.edu.cn}

\author{Donghao Zhu}
\authornote{The authors are listed alphabetically by family name. Donghao Zhu
is the lead and primary contributor to this work.}
\orcid{0000-0002-1889-3076}
\affiliation{%
  \department{Institute of Business Sciences}
  \institution{University of Tsukuba}
  \streetaddress{3-29-1 Otsuka}
  \city{Bunkyo-ku}
  \state{Tokyo}
  \postcode{112-0012}
  \country{Japan}}
\affiliation{%
  \institution{University of Tokyo Market Design Center}
  \streetaddress{7-3-1 Hongo}
  \city{Bunkyo-ku}
  \state{Tokyo}
  \postcode{113-0033}
  \country{Japan}}
\email{zhu.donghao.gt@u.tsukuba.ac.jp}

\renewcommand{\shortauthors}{Xu and Zhu}

\begin{abstract}
Agentic buyers can cheaply fork a procurement task into many parallel negotiations, but concurrency is not free: every thread consumes resources, and simultaneous agreements create cancellation and commitment risk. 
We study a one-unit post-order sourcing problem with a single hard-deadline negotiation window, in which a planner jointly chooses the number of seller-facing negotiators and a common procurement price cap. 
The model combines a product-specific acceptance curve with fulfillment loss, per-thread cost, and excess-commitment cost. 
We establish three structural results. 
First, holding the per-thread acceptance target fixed, the marginal value of another negotiator decays geometrically, yielding a conditional concurrency threshold. 
Second, under a convex quantile curve, parallelism substitutes for concession: more concurrent negotiators imply a weakly lower per-thread acceptance target and price cap. 
Third, when prices are more dispersed, Agentic buyers benefit by searching harder for bargains, but suffer when they instead try to guarantee procurement by offering higher prices.
We operationalize these results in the \emph{Concurrency-Aware Negotiation Optimizer} (\CANO{}), a deterministic optimizer that jointly determines the optimal negotiation concurrency and procurement price cap for an agentic procurement system. 
Across different analytic market configurations and extensive Monte Carlo, finite-data, non-Gaussian, and correlated-seller stress tests, \CANO{} consistently outperforms common heuristic policies while validating the predicted structural properties.

\end{abstract}

\ccsdesc[500]{Computing methodologies~Multi-agent systems}
\ccsdesc[300]{Theory of computation~Algorithmic game theory and mechanism design}
\ccsdesc[300]{Applied computing~Online shopping}

\keywords{automated negotiation, agentic commerce, concurrent negotiation,
procurement, multi-agent orchestration, reservation price}

\begin{document}
\maketitle

\section{Introduction}
Large language model (LLM) agents are beginning to search, bargain, and transact on behalf of users. 
A distinctive capability of such systems is \textit{cheap replication}: a single procurement task can be delegated to multiple seller-facing subagents in parallel. 
Replication increases the chance of finding an acceptable seller while reducing the concession required in any individual negotiation. 
However, simply contacting every seller is rarely optimal.
Each additional negotiation thread consumes tokens, time, and seller attention, while near-simultaneous agreements may leave the buyer with multiple commitments for a single purchase.

This trade-off has attracted attention in both classical automated negotiation and recent language-agent systems.
Prior work on concurrent negotiation studies coordination, tentative agreements, and decommitment~\cite{nguyen2005commitments,an2011strategic,bagga2020deep}, while recent benchmarks evaluate language agents in multi-seller markets with parallel execution~\cite{zhu2025risky,liu2026agenticpay}.
Nevertheless, existing approaches typically treat the number of negotiation threads or the buyer's reservation price as fixed inputs, or learn an end-to-end dialogue policy. 
They do not isolate the fundamental orchestration decision that arises before any dialogue begins: 
\textit{How many negotiation agents should be launched, and what procurement price cap should they share?}

Those controls are coupled. Holding a cap fixed while adding threads improves fulfillment but also increases the chance of duplicate acceptance. 
Holding concurrency fixed while raising the cap improves each thread's success but directly reduces transaction surplus.
Optimizing either control in isolation can therefore recommend a high deal rate and still destroy value. 
The coupling is especially consequential for language agents because replication is cheap enough to make large \(m\) operationally plausible, while their seller-facing actions remain difficult to coordinate after several conversations converge at nearly the same time.

We answer this question for one-unit post-order procurement. After accepting a customer order, a platform observes \(N\) eligible sellers and an estimated offer-acceptance curve \(F\). 
It chooses \(m\leq N\) concurrent negotiators and a common terminal cap \(b\). 
All contacted sellers must be launched before any terminal response arrives: 
The delivery service-level agreement leaves time for one negotiation window, but not for a second serial attempt. 
A seller accepts with probability \(F(b)\).
Failure incurs loss \(L\), each thread costs \(\kappa\), and each acceptance beyond the one fulfilled agreement incurs residual commitment cost \(\delta\).
The last term captures races that remain even with a commitment manager:
cancellation work, seller harm, reputational loss, or a legally binding tentative action.

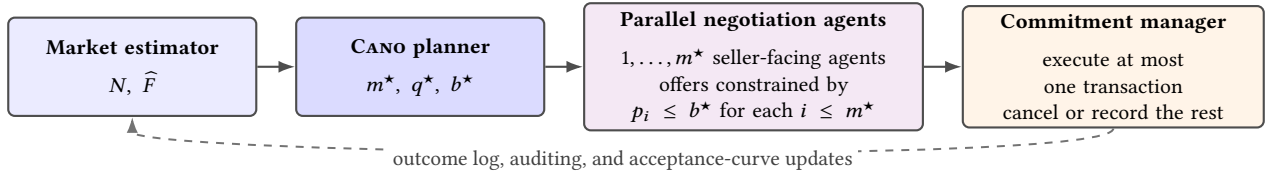
\begin{figure*}[t]
\centering
\begin{tikzpicture}[
    font=\small,
    >=Latex,
    node distance=5mm,
    block/.style={
        draw=black!70,
        line width=0.8pt,
        rounded corners=2.5pt,
        align=center,
        minimum height=13mm,
        inner xsep=4pt,
        inner ysep=3pt
    },
    estimator/.style={
        block,
        fill=blue!8,
        text width=30mm
    },
    planner/.style={
        block,
        fill=blue!14,
        text width=30mm
    },
    agents/.style={
        block,
        fill=violet!9,
        text width=42mm
    },
    manager/.style={
        block,
        fill=orange!9,
        text width=37mm
    },
    flow/.style={
        -{Latex[length=2.2mm,width=1.6mm]},
        line width=0.8pt,
        draw=black!70
    },
    feedback/.style={
        -{Latex[length=2.1mm,width=1.5mm]},
        dashed,
        line width=0.75pt,
        draw=black!55
    },
    feedbacklabel/.style={
        font=\small,
        fill=white,
        inner sep=2pt,
        text=black!75,
        align=center
    }
]


\node[estimator] (estimator) {
    \textbf{Market estimator}\\[1.5mm]
    \(N,\ \widehat F\)
};

\node[planner, right=of estimator] (planner) {
    \textbf{\CANO{} planner}\\[1.5mm]
    \(m^\star,\ q^\star,\ b^\star\)
};

\node[agents, right=of planner] (agents) {
    \textbf{Parallel negotiation agents}\\[1.5mm]
    \(1,\ldots,m^\star\) seller-facing agents\\
    offers constrained by \(p_{i}\leq b^\star\) for each $i \leq m^\star$
};

\node[manager, right=of agents] (manager) {
    \textbf{Commitment manager}\\[1.5mm]
    execute at most one transaction\\
    cancel or record the rest
};


\draw[flow] (estimator.east) -- (planner.west);
\draw[flow] (planner.east) -- (agents.west);
\draw[flow] (agents.east) -- (manager.west);


\draw[feedback]
    (manager.south)
    .. controls
        ($(manager.south)+(0,-5mm)$)
        and
        ($(estimator.south)+(0,-5mm)$)
    ..
    (estimator.south);

\node[feedbacklabel]
    at ($(estimator.south)!0.5!(manager.south)+(0,-5mm)$)
    {outcome log, auditing, and acceptance-curve updates};

\end{tikzpicture}

\caption{\CANO{} as the economic control layer of an agentic procurement
system. The market estimator supplies the eligible-seller count \(N\) and
estimated acceptance curve \(\widehat F\). The planner selects concurrency
\(m^\star\), acceptance target \(q^\star\), and procurement cap \(b^\star\).
Seller-facing agents negotiate in parallel subject to \(p_{it}\leq b^\star\),
and the commitment manager executes at most one binding transaction. Logged
outcomes support auditing and future acceptance-curve updates.}

\Description{A horizontal architecture diagram with four rounded blocks:
market estimator, CANO planner, parallel negotiation agents, and commitment
manager. Solid arrows indicate the forward control flow. A dashed curved arrow
returns outcome logs, audit information, and acceptance-curve updates from the
commitment manager to the market estimator.}
\label{fig:architecture}
\end{figure*}

Our contribution is a small, auditable control problem rather than another end-to-end bargaining policy.
\begin{itemize}
  \item 
  We formulate concurrency and concession jointly through a single expected-profit objective. The objective separates fulfillment value, language-agent operating cost, and excess-commitment exposure.
  \item 
  We characterize a fixed-target concurrency threshold, prove that parallelism substitutes for concession under explicit regularity conditions, and identify a sign reversal in the value of price dispersion. 
  \item 
  We implement \CANO{} and evaluate it over a complete
  \GridScenarioCount{}-configuration factorial sweep. 
  The main comparisons are analytic rather than sample-contingent; one million simulated orders validate
  the expectation, and additional experiments study estimation error, non-Gaussian markets, and correlated seller prices.
\end{itemize}
The results reveal two practical insights. 
First, as the number of sellers increases, Agentic buyers can rely more on parallel search and less on raising prices to secure procurement. 
Second, greater price dispersion creates more opportunities to find bargains, but also makes guaranteed procurement more expensive.


\section{Background and System Framework}

\subsection{Related Work}
\paragraph{Automated and concurrent negotiation.}
Automated negotiation has long combined game-theoretic, heuristic, and argumentation-based approaches~\cite{jennings2001automated}. 
Parallel protocols and commitment models allow one agent to bargain with several counterparties while controlling delay and decommitment~\cite{aknine2004extended,nguyen2005commitments}.
An et al.~\cite{an2011strategic} coordinate tentative agreements and concessions for multiple resources, and Bagga et al.~\cite{bagga2020deep} learn one-to-many policies with deep reinforcement learning. 
The recent taxonomy of Florijn et al.~\cite{florijn2025survey} emphasizes that one-to-many settings differ principally in how negotiation outcomes interact. 
Our setting is the unit-demand case: successes are substitutes, and at most one can be retained.
Unlike dialogue-level strategies, we obtain a product-aware ex ante rule for both the number of dialogues and their shared reservation cap.

\paragraph{Language-mediated bargaining.}
End-to-end negotiation dialogue models jointly learn strategy and language~\cite{lewis2017deal}; modular models instead decouple a strategic act from its linguistic realization~\cite{he2018decoupling}. 
This separation has become especially relevant for LLM agents. 
NegotiationArena, buyer-side bargaining evaluations, and self-play studies document strategic variation and irrational behavior~\cite{bianchi2024well,xia2024bargaining,fu2023improving}.
Consumer-market experiments report overpayment and unfavorable
transactions~\cite{zhu2025risky}, while verifiable-reward training improves budget adherence and surplus extraction~\cite{liu2026rlvr,liu2026strategic}.
AgenticPay provides 111 tasks from bilateral to many-to-many negotiation and directly contrasts sequential and parallel modes~\cite{liu2026agenticpay}.
\CANO{} is complementary: it is an outer economic controller, not a language policy or benchmark. In particular, its cap can be enforced after an LLM proposes a structured price action.

\subsection{Positioning}
Our unit of analysis is an orchestration decision, rather than an utterance, dialogue trajectory, or learned bargaining equilibrium. 
This creates two differences from the closest systems work. 
First, the policy is chosen \emph{before} seller-facing text is generated and is expressed in quantities that an execution layer can enforce: a thread count and a terminal cap. 
Second, the objective values fulfillment, price, computation, and duplicate commitments jointly; agreement rate alone is not the target. 
The theory does not assert that one language model persuades better than another. 
It specifies the \textit{economic authority} within which any such model may act and supplies comparative statics that dialogue benchmarks can test.

Figure~\ref{fig:architecture} illustrates the overall workflow of \CANO{}. 
The market estimator learns the eligible seller count \(N\) and acceptance curve \(\widehat F\) from historical offer--response data.
Based on these estimates, the planner jointly determines the optimal concurrency \(m^\star\), acceptance target \(q^\star\), and procurement price cap \(b^\star\). 
The selected seller-facing agents negotiate in parallel under the shared price cap, while the commitment manager executes at most one transaction and records the remaining commitments.
The resulting outcome log is fed back to the market estimator to improve future acceptance-curve estimation.

\subsection{System Architecture}
At planning time, the estimator supplies a product-state snapshot, an eligible seller set, and an estimated acceptance curve \(F\). 
The planner uses the corresponding quantile function $C(q)=F^{-1}(q)$ to translate a target per-thread acceptance probability into the corresponding procurement price cap. 
It then either returns ``cancel'' or a tuple \((m^\star,q^\star,b^\star)\) together with a selected set of \(m^\star\) sellers.
During execution, each language agent may propose negotiation actions, but every proposed action is validated by the execution layer before being dispatched. 
In particular, the execution layer enforces
\begin{equation*}
p_{i}\leq b^\star
\quad\text{and}\quad
\sum_{i=1}^{m^\star}x_i \leq 1,
\end{equation*}
where \(x_i\in\{0,1\}\) is a binary commitment indicator that equals 1 if seller \(i\) is selected for the final binding transaction, and 0 otherwise.
The parameter \(\delta\) represents the residual cost incurred when simultaneous negotiations cannot be perfectly coordinated because of external latency, seller-side finalization, or legally binding tentative acceptance.
It models unavoidable implementation risk rather than relaxing the execution constraints.

The outcome log records the estimator version, planned tuple, offers, acceptances, rejected actions, final commitment, and cancellations. 
These records support off-policy recalibration of \(F\), cap-violation audits, and human review without exposing the language model to unilateral authority over payment. 
Thus Figure~\ref{fig:architecture} separates a statistical component that may be uncertain, an optimization component that is deterministic given its inputs, and an execution component that is safety critical.
The architecture is presented to clarify the system context rather than to optimize every component. 
In particular, this paper focuses exclusively on the economic control layer, the planner that determines negotiation concurrency and procurement price caps, while estimator design, language policy, and execution engineering are outside the scope of this work.


\section{Procurement Model}

\begin{table}[!t]
  \caption{Core notation and operational interpretation.}
  \label{tab:notation}
  \centering
  \small
  \begin{tabular}{@{}cl@{}}
    \toprule
    Symbol & Meaning \\
    \midrule
    \(N\) & eligible sellers (market thickness) \\
    \(m\) & concurrently launched negotiators \\
    \(b=C(q)\) & common terminal procurement cap \\
    \(V\) & net value of a fulfilled order \\
    \(L\) & loss from cancellation or failed fulfillment \\
    \(\kappa\) & cost per launched negotiation thread \\
    \(\delta\) & cost per acceptance beyond the first \\
    \bottomrule
  \end{tabular}
\end{table}

\subsection{Market and Decisions}

One accepted customer order yields net value \(V>0\) if fulfilled, after non-procurement costs. 
There are \(N\) sellers that already satisfy quality and delivery filters. 
Seller \(i\)'s minimum acceptable terminal price is \(R_i\sim F\), independently and identically distributed conditional on the product state. 
The planner starts \(m\in\{1,\ldots,N\}\) threads and assigns the same terminal procurement cap \(b\) to every thread. 
It may instead cancel immediately, denoted \(m=0\), receiving payoff \(-L\).
We remark that, if procurement is expected to be uneconomical, the platform may immediately cancel the order and refund the customer instead of initiating negotiation.
The cap is a \emph{walk-away price}, not an opening offer.
The terminal-offer abstraction charges \(b\) whenever at least one seller accepts. 
It is conservative when a multi-round protocol sometimes closes below its cap~\cite{baarslag2016learning}.

A seller accepts if and only if \(R_i\leq b < V\). 
We write $q=F(b)$ and $C(q)=F^{-1}(q)$ where \(q\) is per-thread acceptance probability and \(C\) is the reservation-price quantile curve. 
``Price dispersion'' below always refers to dispersion in the latent distribution of seller reservation prices \(R_i\), rather than dispersion in posted prices.
For the structural results, \(F\) is continuous and strictly increasing on the price interval containing the active policies, so the displayed inverse is well defined. 
The implementation can instead consume a bounded empirical quantile and uses global numerical search rather than the derivative results.

\textit{Remark (timing and information).}
The customer order is already accepted when sourcing begins. The platform first applies non-price eligibility filters and freezes a product-state estimate of \(F\); 
It then chooses \(m,b\), launches the threads simultaneously, observes terminal acceptances at the end of the sourcing window, and retains at most one transaction. 
There is no opportunity to launch a replacement thread after observing failure without missing the fulfillment deadline. 
Outcomes update future estimates but do not change \(F\) within the current order. 
After conditioning on observable seller and product characteristics through eligibility filtering and market estimation, the remaining reservation prices are assumed to be i.i.d.
Shared unobserved shocks violate it and are evaluated separately in Section~\ref{sec:stress}.

We restrict attention to a common cap. Exchangeability motivates this
parsimonious choice, and it reduces the execution interface to one auditable
scalar, but we do not claim it is optimal among arbitrary asymmetric
mechanisms. If seller-specific features support distinct conditional curves
\(F_i\), the planner should choose seller-specific caps and a subset jointly;
that heterogeneous assignment problem is outside the structural claims here.
Likewise, \(F\) describes the terminal response to a cap under a fixed
interaction protocol. It is not identified by posted prices, and it may change
if the dialogue policy, seller population, or product state changes.

\subsection{Expected Objective}

For \(m\geq1\), the number of acceptances is \(K\sim\operatorname{Binomial}(m,q)\). 
Hence
\[
S_m(q)=\Prob(K\geq1)=1-(1-q)^m
\]
and
\begin{equation}
D_m(q) = \E[(K-1)^+]=mq-S_m(q).\label{eq:excess-acceptance}
\end{equation}
When \(K=0\), payoff is \(-L-\kappa m\). 
When \(K\geq1\), the buyer keeps one agreement and receives \(V-b-\kappa m-\delta(K-1)\). 
Taking expectations gives
\begin{align}
\Pi(m,b)
={}& 
\underbrace{(V-b)S_m(F(b))}_{\text{Expected fulfilled-order value}}
-
\underbrace{L[1-S_m(F(b))]}_{\text{Expected fulfillment loss}}
\nonumber\\
&
-
\underbrace{\kappa m}_{\text{Thread cost}}
-
\underbrace{\delta\{mF(b)-S_m(F(b))\}}_{\text{Expected excess-commitment cost}}.
\label{eq:profit-b}
\end{align}
The objective~\eqref{eq:profit-b} consists of four economically meaningful components.
The first term is the expected surplus generated when procurement succeeds, namely the net order value multiplied by the fulfillment probability.
The second term is the expected loss from procurement failure.
The third term captures the operating cost of launching negotiation threads.
The final term represents the expected cost of excess commitments caused by multiple simultaneous acceptances.
In acceptance space this becomes the central objective:
\begin{equation}
\boxed{\Pi(m,q)=
[V+L+\delta-C(q)]S_m(q)-L-\kappa m-\delta mq.}
\label{eq:profit-q}
\end{equation}

In addition to profit and fulfillment, we report two complementary measures of commitment risk.
The expected number of acceptances beyond the first is
\(D_m(q)\), defined in~\eqref{eq:excess-acceptance}.
The probability that at least two threads accept is
\begin{equation*}
B_m(q)=1-(1-q)^m-mq(1-q)^{m-1}.
\end{equation*}
\(D_m\) prices the linear exposure in the objective, whereas \(B_m\) is easier
to interpret as the frequency of orders requiring decommitment handling. Both
are zero for \(m=1\), but neither is controlled by fulfillment probability
alone.

The planner maximizes \(\Pi\) over \(m\in\{1,\ldots,N\}\) and \(q\in[0,1]\),
then compares the best procurement policy with immediate cancellation
\(\Pi(0)=-L\). Appendix~\ref{app:derivation} gives the full derivation and
boundary conventions.



\section{Structural Results and \CANO{}}

\subsection{A Threshold for Concurrency}

Fix \(q\). The value of adding a \((m+1)\)-st thread is
\begin{equation}
\Delta_m(q)=[V+L+\delta-C(q)]q(1-q)^m-\kappa-\delta q.
\label{eq:marginal}
\end{equation}
\begin{proposition}[Concurrency threshold]
\label{prop:threshold}
For fixed \(q\in(0,1)\):
\begin{enumerate}
    \item If \(V+L+\delta-C(q)>0\), then $\Delta_{m+1}(q)<\Delta_m(q)$ for all $m \geq 1$.
    \item If \(V+L+\delta-C(q)\leq0\), then $\Delta_m(q)\leq0$ for all $m \geq 1$.
    \item With ties resolved in favor of fewer threads, the optimizer of the fixed-\(q\) problem is 
    \begin{equation*}
        m_q^\star = \min\left\{ m\in\{1,\ldots,N\}: m=N \text{ or } \Delta_m(q)\leq0 \right\}.
    \end{equation*}
\end{enumerate}
\end{proposition}
\noindent
Proposition~\ref{prop:threshold} establishes a threshold structure for concurrency. 
For a fixed per-thread acceptance target, the planner continues adding negotiators while the next thread has positive incremental value and stops once this value becomes non-positive. 
The comparative statics follow immediately from
\eqref{eq:marginal}. 
Increasing \(V\) or \(L\) raises the incremental value of
every additional thread, whereas increasing \(\kappa\) or \(\delta\) lowers it.
Consequently, the threshold weakly increases with \(V\) and \(L\), and weakly decreases with \(\kappa\) and \(\delta\).
The intuition is event-based. 
Thread \(m+1\) creates new fulfillment value only when all first \(m\) threads fail, an event with probability \((1-q)^m\).
Operating and excess-commitment costs do not enjoy this geometric discount.
The complete proof is in Appendix~\ref{app:proof-threshold}.

The threshold is directly computable without enumerating all subsets. 
If \(\kappa+\delta q>0\), the stopping point is the
smallest \(m\geq1\) such that
\begin{equation}
  (V+L+\delta-C(q))q(1-q)^m\leq \kappa+\delta q.
  \label{eq:thread-stop}
\end{equation}
This expression separates a \emph{scarcity term}, \((V+L+\delta-C(q))q(1-q)^m\), from a
nonvanishing launch-and-commitment cost. It also shows why a large seller pool
does not imply contacting every seller: \(N\) enlarges the feasible set, but it
does not slow the geometric decay of the next thread's contribution. If
\(\kappa=\delta=0\), every finite marginal is positive and the fixed-\(q\)
subproblem selects \(N\).





\subsection{Parallelism Substitutes for Concession}

For fixed \(m\), define
\[
A_m(q)=\frac{1-(1-q)^m}{m(1-q)^{m-1}}
=\frac{q}{m}\sum_{j=0}^{m-1}(1-q)^{-j}.
\]
We state the substitution result under conditions that rule out spurious
nonconvex roots.
\begin{assumption}[Regular quantile]
\label{ass:regular}
Let \(I=\{q\in(0,1):C(q)\leq V+L+\delta\}\). On \(I\), \(C\) is twice continuously differentiable, \(C'(q)>0\), and \(C''(q)\geq0\). 
For every fixed \(m\) under comparison, every global maximizer over \(q\in[0,1]\) lies in the
interior of \(I\).
\end{assumption}
\noindent
Assumption~\ref{ass:regular} imposes mild regularity conditions on the reservation-price quantile curve. 
Monotonicity (\(C'(q)>0\)) reflects the natural requirement that achieving a higher acceptance probability requires a weakly higher procurement cap. 
Convexity (\(C''(q)\ge0\)) rules out pathological quantile shapes and ensures that progressively higher acceptance targets become increasingly expensive. 
Finally, restricting the optimizer to the interior of \(I\) excludes degenerate corner solutions, such as always rejecting or always accepting, and ensures that the optimum is fully characterized by the first-order condition, allowing the comparative-statics analysis to focus on economically meaningful operating regimes.

\begin{theorem}[Parallelism--concession substitution]
\label{thm:substitution}
Under Assumption~\ref{ass:regular}, the fixed-\(m\) optimizer \(q_m^\star\) is
unique and satisfies
\begin{equation}
C(q)+C'(q)A_m(q)
+\delta\!\left[(1-q)^{-(m-1)}-1\right]=V+L.
\label{eq:foc}
\end{equation}
Moreover,
\[
q_{m+1}^\star\leq q_m^\star \ \text{ and }\ 
C(q_{m+1}^\star)\leq C(q_m^\star).
\]
\end{theorem}

Theorem~\ref{thm:substitution} shows that the optimal bargaining policy depends on the available degree of parallelism. 
As more negotiation agents (larger $m$) are launched
in parallel, the planner optimally lowers the target per-thread acceptance probability, which in turn implies a lower procurement cap. 
Consequently, parallelism substitutes for concession: additional negotiation threads allow each individual agent to bargain more aggressively rather than relying on higher prices to secure procurement.
Therefore, an agentic buyer should adapt its bargaining policy to the available degree of parallelism instead of replicating one fixed strategy across different market conditions.
Conversely, when API or seller-contact cost raises \(\kappa\), the joint solution typically uses fewer threads and compensates with a higher cap. 
The proof in Appendix~\ref{app:proof-substitution} shows that both \(A_m(q)\) and the commitment term in~\eqref{eq:foc} grow with \(m\), forcing the unique root leftward.

Equation~\eqref{eq:foc} is a marginal balance condition: the planner increases the procurement cap until the marginal gain from higher procurement reliability is exactly offset by the marginal procurement and duplicate-commitment costs.
As more negotiation agents are launched in parallel, overall procurement becomes less dependent on any single negotiation thread, reducing the need to buy reliability through higher prices. 
Consequently, parallelism substitutes for concession: more parallel agents imply a lower optimal acceptance target
and procurement cap for each individual negotiator. 
The thread cost \(\kappa\) does not appear in the first-order condition because it is independent of the procurement cap for fixed \(m\). 
Instead, it affects the outer optimization that determines the optimal number of negotiation threads. 
Managerially, the theorem suggests that, when parallelism is inexpensive, agentic buyers should invest computational resources in broader parallel search rather than higher procurement prices.



\subsection{When Dispersion Helps}
Consider a mean-preserving symmetric location-scale family, in which the mean reservation price remains fixed while the distribution becomes more or less dispersed symmetrically around the mean:
\[
C_\sigma(q)=\mu+\sigma z(q)
\ \text{ and }\ 
z(1-q)=-z(q),
\]
where \(\sigma\) controls the degree of price dispersion and symmetry implies that reservation prices spread equally above and below the mean.
So \(z(q)\) is negative below the median and positive above it. The uniform
family used for the main experiment has
\[
z(q)=\sqrt{3}(2q-1)
\ \text{ and}\ 
R_i\sim U\left[\mu-\sqrt{3}\sigma,\mu+\sqrt{3}\sigma\right].
\]

\begin{theorem}[Dispersion sign reversal]
\label{thm:dispersion}
Fix \(m\). Suppose its maximizing target is unique and interior. Let
\(\Pi_m^\star(\sigma)\) denote the resulting maximized profit. At differentiability
points,
\[
\frac{d\Pi_m^\star(\sigma)}{d\sigma}
=-z(q_m^\star)S_m(q_m^\star).
\]
Hence dispersion raises optimal value when \(q_m^\star<1/2\), lowers it when
\(q_m^\star>1/2\), and is locally neutral at \(q_m^\star=1/2\).
\end{theorem}

Parallel search makes the lower tail usable. 
In a thick market, where many eligible sellers are available, the buyer can achieve high overall fulfillment even when each individual negotiation has a relatively low acceptance probability (\(q_m^\star<1/2\)).
Consequently, spreading reservation prices creates more opportunities to find low-price sellers.
By contrast, in a thin market, where only a few eligible sellers are available, or when fulfillment reliability is critical, the planner must keep the target acceptance probability above the median.
Although Theorem~\ref{thm:dispersion} is stated for a fixed concurrency level \(m\), the same conclusion extends piecewise to the joint optimization problem. 
Specifically, whenever the optimal concurrency \(m^\star(\sigma)\) remains constant over an interval of \(\sigma\), the value function satisfies $\Pi^\star(\sigma) = \Pi_{m^\star}^\star(\sigma)$ and the dispersion result applies directly. 
Non-differentiability may occur only at switching points where the optimizer changes from one concurrency level to another, i.e., \(m^\star(\sigma^-)\neq m^\star(\sigma^+)\). 
The complete envelope argument is provided in Appendix~\ref{app:proof-dispersion}.

The result is deliberately local to a location--scale family. 
It does not say that every mean-preserving spread is beneficial below the median, nor that an increase in observed listing-price variance has this effect. 
The operative object is the estimated terminal acceptance quantile. 
The sign result is useful because it converts a distributional change into a testable policy boundary: 
After re-optimization, the active target \(q^\star=.5\) separates a lower-tail-search regime from a reliability-through-concession regime.



\subsection{\CANO{} Optimizer}

\CANO{} enumerates \(m=1,\ldots,N\), solves the one-dimensional fixed-\(m\) problem, evaluates~\eqref{eq:profit-q}, and compares the best result with \(-L\). 
Under Assumption~\ref{ass:regular}, the left side of~\eqref{eq:foc} is strictly increasing, so bisection with boundary checks returns a globally optimal policy to tolerance \(\varepsilon\) using \(O(N\log(1/\varepsilon))\) evaluations of the quantile function and its derivative.
Empirical acceptance curves and mixtures need not satisfy the assumption. 
Our implementation therefore uses a bounded global scan in \(q\), followed by golden-section refinement around the best cell.
Unlike derivative-based root-finding methods, this implementation does not rely on a favorable initialization and therefore remains robust for empirical acceptance curves.
This version costs \(O(NG+N\log(1/\varepsilon))\) where \(G\) is the number of grid points in the initial coarse search. 
The first term corresponds to the global scan, while the second is due to the subsequent golden-section refinement.
Complete implementation details, including pseudocode, correctness, boundary handling, and numerical validation, are provided in Appendix~\ref{app:algorithm}.


\section{Computational Evaluation}

\subsection{Design}

We ask three questions: whether the numerical optimizer reproduces the
structural results; how costly common restricted policies are; and how sensitive
the planner is to estimating \(F\) or misspecifying seller dependence.

\paragraph{Parameters and policies.}
We normalize \(\mu=1\). The complete analytic factorial crosses
\[
\begin{aligned}
N&\in\{1,2,5,10,20,50\},&
\sigma&\in\{.05,.15,.30,.45\},\\
V-\mu&\in\{.1,.3,.5\},&
L&\in\{0,.1,.3\},\\
\kappa&\in\{.001,.01,.03\},&
\delta&\in\{0,.02,.1\},
\end{aligned}
\]
for \(\GridScenarioCount{}\) configurations. We compare joint \CANO{} with:
(i)~\textsc{Single-agent} Optimal (\(m=1\), optimize \(q\)), representing the traditional one-to-one negotiation paradigm without parallelism;
(ii)~\textsc{All-parallel} (\(m=N\), optimize \(q\)), representing the strategy of launching every available negotiation agent;
(iii)~\textsc{Mean-cap} (\(b=\mu\), optimize \(m\)), representing a simple market-heuristic pricing rule with a fixed procurement cap at the market mean; and
(iv)~\textsc{Fixed-\(q\)} (\(q=.8\) per thread, optimize \(m\)), representing a service-level heuristic that fixes a conventional high per-thread acceptance target while adjusting only concurrency.
Every policy may immediately cancel. 
Because these baselines restrict one or both decision dimensions, they quantify the value of joint optimization.
For each market configuration, we define the regret of a restricted policy as
\[
R_{\mathrm{policy}}
=
\max_{\substack{m\in\{0,\ldots,N\}\\ q\in[0,1]}}
\Pi_{\CANO}(m,q)
-
\Pi_{\mathrm{policy}},
\]
where $\mathrm{policy} \in \{ 
\textsc{Single-agent},
\textsc{All-parallel},
\textsc{Mean-cap},
\textsc{Fixed-}q
\}
$ and $\Pi_{\CANO}(m,q)$ denotes the expected profit achieved by the proposed joint optimizer.
Regret therefore measures the opportunity cost of restricting the feasible decision space, and the informative outcome is its magnitude and location rather than the fact that it is nonnegative.
We additionally report the ratio of aggregate surplus above \(-L\), the share of cells within one percent of the oracle surplus (with a \(0.01\) scale floor near zero), and cancellation frequency. 
No confidence interval is attached to the factorial sweep because it exhausts the declared design rather than sampling configurations.

Main comparisons use analytic expected values and therefore have no sampling interval. 
Reference runs use 2,049 \(q\)-values before refinement; 
The factorial and regime-map runs use 257, estimation and dependence runs use 513, and distributional and sensitivity runs use 1,025. 
These resolutions were selected through preliminary convergence checks, balancing numerical accuracy against computational cost; denser grids produced negligible changes in the resulting objective values while requiring substantially more evaluations.
Empirical-quantile runs also include every interpolation knot. 
Sixteen unit tests cover the expectation, marginal identity, first-order condition, monotonicity, dispersion derivative, distribution inversion, cancellation, correlated extension, and agreement with a 100,001-point exhaustive grid. 
A one-million-order Monte Carlo run uses seed 20,260,803. Further details are in Appendix~\ref{app:experiments}.

\subsection{Structural and Policy Results}
\begin{figure}[t]
  \centering
  \includegraphics[width=0.5\textwidth]{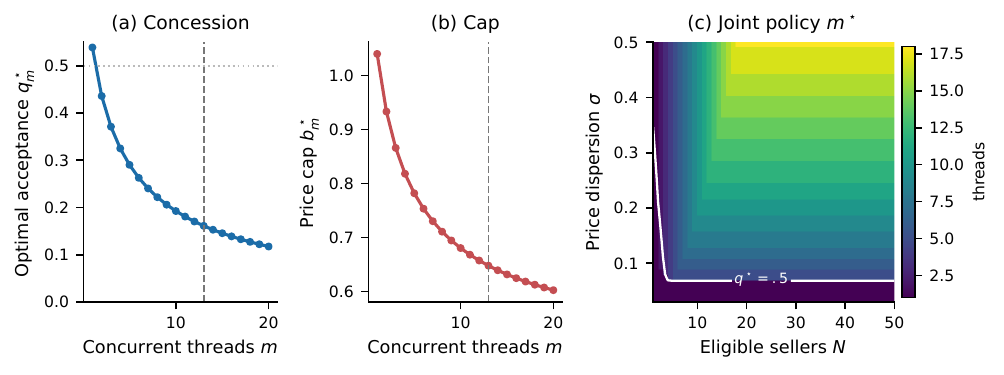}
  \caption{Structural predictions for \(V=1.4,\mu=1,L=.2,\kappa=.01,\delta=.05\).
  (a--b) With \(N=20,\sigma=.30\), the fixed-\(m\) acceptance target and cap
  fall as concurrency grows; the dashed line marks the joint optimum.
  (c) The joint policy over market thickness and dispersion. The white contour
  marks \(q^\star=.5\), the sign boundary in Theorem~\ref{thm:dispersion}.}
  \Description{Three panels. The first plots an acceptance target decreasing
  with concurrent threads. The second plots a decreasing price cap. The third
  is a heat map in seller count and price dispersion, with higher thread counts
  in thicker and more dispersed markets and a white median contour.}
  \label{fig:theory}
\end{figure}

\paragraph{Policy adaptation across market regimes.}
Figure~\ref{fig:theory} (a) and (b) sweep the candidate concurrency level \(m=1,\ldots,N\). 
At each value of \(m\), the planner holds \(m\) fixed and solves the one-dimensional problem $q_m^\star\in\arg\max\limits_{q\in[0,1]}\Pi(m,q)$.
Panel~(a) plots the resulting conditional optimum \(q_m^\star\), whereas panel~(b) expresses the same solution in price units through $b_m^\star=C(q_m^\star)$.
Thus, panel~(b) is the price-space representation of panel~(a), rather than the result of a separate optimization over \(b\). 
The vertical dashed line marks the jointly optimal concurrency \(m^\star\), obtained by comparing \(\Pi(m,q_m^\star)\) across all candidate thread counts and with immediate cancellation.
Panel~(c) instead varies the market environment. 
Its horizontal axis is the number of eligible sellers \(N\), and its vertical axis is reservation-price dispersion \(\sigma\). 
At each point \((N,\sigma)\), \CANO{} jointly re-optimizes \(m\) and \(q\); the color reports the resulting \(m^\star\), with darker colors indicating fewer threads and brighter colors indicating more. 
The stepwise color bands reflect the discreteness of \(m^\star\). 
The white contour is the set of market configurations for which the jointly optimized target satisfies \(q^\star=.5\). 
The higher-dispersion side of the contour corresponds to \(q^\star<.5\), whereas the lower-dispersion side corresponds to \(q^\star>.5\). 
The point \((N,\sigma)=(20,.30)\) in panel~(c) is exactly the reference market shown in panels~(a) and~(b): it yields $(m^\star,q^\star,b^\star)
=(\ReferenceMStar,\ReferenceQStar,\ReferenceCapStar)$ and expected profit \(\ReferenceProfit\).
Although the optimal per-thread acceptance target falls from \(q_1^\star=.539\) to \(q^\star=\ReferenceQStar\), aggregate fulfillment rises to $S_{m^\star}(q^\star) =1-(1-q^\star)^{m^\star} \approx .898$.

Figure~\ref{fig:theory} translates the structural results into an operating map. 
Panels~(a) and~(b) show how the optimized per-thread acceptance target and procurement cap adjust as concurrency increases, while panel~(c) identifies the market conditions under which broader parallel search becomes valuable. 
At low dispersion, the optimizer uses only a few relatively reliable threads because there is little lower-tail value to exploit. 
As \(N\) and \(\sigma\) increase, optimal concurrency generally rises but remains well below \(N\); 
The downward-sloping \(q^\star=.5\) contour further indicates that thicker markets can enter the lower-tail-search regime at lower levels of dispersion. 
Managerially, eligible sellers should therefore be treated as an option pool rather than as a launch target: 
Broader search should be paired with tighter price authority, and the buyer should choose between reliability through concession and reliability through parallel lower-tail search according to market thickness and the estimated reservation-price distribution.

\begin{figure}[t]
  \centering
  \includegraphics[width=0.5\textwidth]{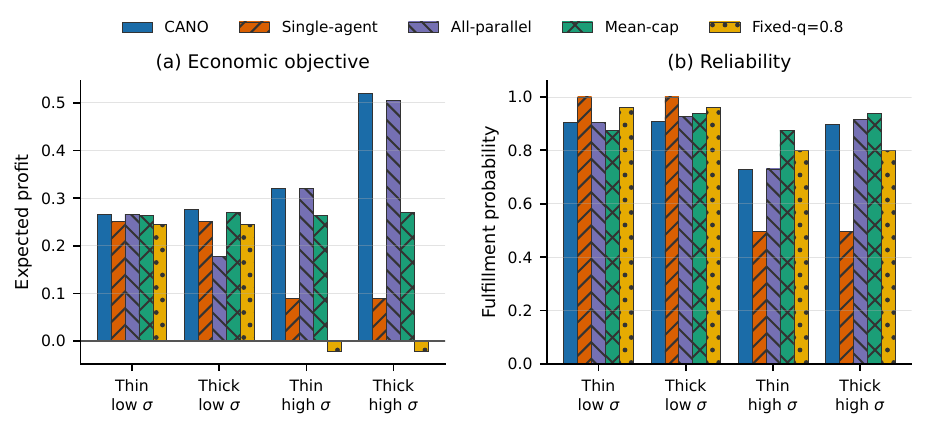}
  \caption{Representative markets with \(N\in\{3,20\}\) and
  \(\sigma\in\{.08,.35\}\), holding \(V=1.4,L=.2,\kappa=.01,\delta=.05\).
  Joint optimization matters most in dispersed markets. Higher reliability is
  not free: Mean-cap and Fixed-\(q\) can fulfill more often while earning less.}
  \Description{Two grouped bar charts over four market regimes. The left shows
  expected profit for five policies and the right shows fulfillment
  probability. CANO has the highest profit, especially in dispersed markets;
  some more reliable policies have lower profit.}
  \label{fig:performance}
\end{figure}

\paragraph{Profit--reliability trade-off.}
Figure~\ref{fig:performance} compares expected profit in panel~(a) and system-level fulfillment probability in panel~(b) across four representative \((N,\sigma)\) regimes. 
Policy differences widen substantially as price dispersion increases. 
In the thick dispersed market, \CANO{} earns \(0.519\) at \(m=14,q=.151\), compared with \(0.090\) for Single-agent and \(0.269\) for Mean-cap, even though All-parallel and Mean-cap attain slightly higher fulfillment.
Together, the two panels show that higher reliability may be too costly when achieved through excessive price concession or unnecessary concurrency. 
\CANO{} instead selects the profit-maximizing reliability level by balancing procurement price, fulfillment loss, thread cost, and excess-commitment exposure.


\begin{table*}[t]
  \caption{Full-factorial policy summary over 1,944 analytic configurations.
  Surplus ratio aggregates profit above immediate cancellation. Regret is in
  normalized profit units; ``Near'' is the share within 1\% of \CANO{}'s
  recoverable surplus; ``Cancel'' is the share choosing \(m=0\).}
  \label{tab:factorial-summary}
  \centering
  \small
  \begin{tabular}{@{}lrrrrrr@{}}
    \toprule
    Policy & Surplus ratio & Mean regret & 90th-pct. regret
      & Near (\%) & Cancel (\%) & Mean \(m\) \\
    \midrule
    \CANO{}              & 1.000 & 0.000 & 0.000 & 100.0 & 0.0  & 7.07 \\
    Single-agent         & 0.614 & 0.160 & 0.425 & 28.8  & 0.0  & 1.00 \\
    All-parallel         & 0.857 & 0.059 & 0.234 & 62.8  & 10.9 & 10.46 \\
    Mean-cap             & 0.727 & 0.113 & 0.319 & 8.5   & 0.0  & 2.96 \\
    Fixed-\(q=.8\)       & 0.440 & 0.232 & 0.575 & 3.6   & 25.9 & 1.23 \\
    \bottomrule
  \end{tabular}
\end{table*}

\paragraph{Costs of restricted orchestration.}
Table~\ref{tab:factorial-summary} summarizes performance over the full factorial and reveals two distinct costs of restricted orchestration.
Single-agent, All-parallel, Mean-cap, and Fixed-\(q\) recover \(61.4\%\), \(85.7\%\), \(72.7\%\), and \(44.0\%\), respectively, of \CANO{}'s \emph{recoverable surplus}, defined as aggregate profit above the immediate cancellation payoff \(-L\). 
All-parallel is within one percent of the joint solution in \(62.8\%\) of configurations, but cancels in \(10.9\%\) because launching all \(N\) threads can be worse than stopping, whereas Single-agent is within one percent in only \(28.8\%\). 
The underlying regret patterns clarify why. 
Single-agent leaves lower-tail search unused: its mean regret rises from \(0.016\) at \(\sigma=.05\) to \(0.300\) at \(\sigma=.45\), and from zero at \(N=1\) to \(0.252\) at \(N=50\). 
All-parallel accesses the lower tail but pays for threads it does not need, with mean regret increasing from zero at \(N=1\) to \(0.211\) at \(N=50\). 
The joint policy instead expands concurrency sublinearly: Averaged over all other factors, mean \(m^\star\) rises from \(1.00\) at \(N=1\) to \(17.06\) at \(N=50\), while the corresponding mean target falls from \(q^\star=.610\) to \(.291\). 
Aggregate fulfillment nevertheless increases from \(.610\) to \(.888\), showing that lower per-thread concession and higher system-level reliability can coexist. 
These comparisons quantify the opportunity cost of restricting the orchestration decision space; 
They do not claim that deterministic optimization outperforms an unconstrained learned policy.

\subsection{Estimation and Dependence Stress Tests}
\label{sec:stress}

We next study how estimation error affects planning by replacing the true quantile function with an empirical quantile estimated from \(n\in\{25,50,100,250,1000\}\) historical reservation samples.
For each sample size, we generate 100 independent datasets and evaluate the resulting policy under the true distribution. 
Mean oracle regret decreases from \(0.0686\) (95\% bootstrap interval \([0.0528,0.0860]\)) at \(n=25\) to \(\EstimationLargeRegret\) (\([0.0027,0.0041]\)) at \(n=1000\), while mean acceptance-calibration error falls from \(.059\) to \(.010\). 
Thus, even an estimated acceptance curve is sufficient for good planning, although limited data can still lead to noticeably suboptimal concurrency decisions.
This experiment intentionally represents a favorable estimation setting because it directly observes sellers' latent reservation thresholds. 
In practice, historical negotiation logs contain only offer--response outcomes, so reservation thresholds are interval-censored and may also be affected by the policies that generated the data. 
Estimating \(\widehat F\) in deployment therefore requires additional modeling assumptions. 
Our results isolate only the effects of finite-sample estimation error and integer policy decisions, leaving these identification issues aside.
\begin{table}[t]
  \caption{Finite-data planning for \(N=20,\mu=1,\sigma=.35,V=1.4,L=.2,
  \kappa=.01,\delta=.05\). Each row uses 100 independent histories; intervals
  are percentile bootstrap intervals for mean oracle regret. Calibration is
  mean absolute acceptance error at the selected cap.}
  \label{tab:estimation}
  \centering
  \small
  \begin{tabular}{@{}rrrrr@{}}
    \toprule
    \(n\) & Mean regret & 95\% interval & Mean \(m\) & Calibration \\
    \midrule
      25 & .0686 & [.0528, .0860] & 12.56 & .0591 \\
      50 & .0251 & [.0197, .0309] & 12.95 & .0337 \\
     100 & .0220 & [.0171, .0273] & 13.53 & .0306 \\
     250 & .0092 & [.0070, .0115] & 14.06 & .0173 \\
    1000 & .0033 & [.0027, .0041] & 13.99 & .0095 \\
    \bottomrule
  \end{tabular}
\end{table}
Table~\ref{tab:estimation} summarizes the finite-data results. 
Because each sample size uses independently regenerated histories rather than nested datasets, the bootstrap intervals for \(n=50\) and \(n=100\) overlap, so their small difference should not be interpreted as a monotone improvement.
Nevertheless, increasing the amount of historical data substantially reduces both planning regret and acceptance-calibration error. 
The reported calibration metric is local to the selected procurement cap and therefore does not assess the accuracy of the entire estimated acceptance curve.

Estimation error is relatively modest compared with violating the independence assumption. 
We therefore introduce an exchangeable Gaussian copula while preserving every seller's marginal acceptance distribution \(F\), where \(\rho\) denotes the latent correlation governing shared seller shocks.
Complete model details are provided in Appendix~\ref{app:correlation}. 
When \(\rho=.7\), representing a strongly correlated market, the policy planned under independence sees expected profit fall from \(0.519\) to \(0.042\), whereas a correlation-aware planner recovers profit to \(0.169\) by reducing concurrency from \(m=14\) to \(4\) and increasing the acceptance target from \(q=.151\) to \(.363\). 
As \(\rho\) increases from \(.1\) to \(.7\), the planner progressively selects fewer but more accommodating negotiation threads, showing that positive dependence weakens the diversification benefit of parallel negotiation: although each seller's marginal acceptance probability remains \(q\), fulfillment is no longer \(1-(1-q)^m\).

Finally, the one-million-order Monte Carlo simulation differs from the analytic expected profit by only \MonteCarloError, within two Monte Carlo standard errors, confirming the numerical correctness of the analytical model. 
Appendix~\ref{app:robustness} reports the complete robustness analyses, including alternative reservation-price distributions, copula-based dependence models, parameter sensitivity analyses, and the full performance curves.

\subsection{What the Evaluation Establishes}
The evaluation establishes that the proposed economic controller behaves consistently with its theoretical predictions across the declared design space. 
The numerical optimizer reproduces the structural comparative statics, joint optimization consistently outperforms common restricted policies across market regimes, and the analytical objective closely matches Monte Carlo simulation. 
Additional finite-data, distributional, and dependence experiments demonstrate robustness to estimation error while identifying seller dependence as the principal source of model misspecification. 
These results validate the economic control layer rather than a complete agentic procurement system. 
They do not establish LLM instruction following, stationary seller behavior, or buyer profit as the appropriate social objective. 
Our contribution is an auditable optimization policy and execution boundary that future end-to-end agent evaluations can build upon, independent of any specific language model or API version.



\section{Limitations, Responsible Use, and Conclusion}

\subsection{Limitations}
\textit{The model deliberately isolates a one-unit, single-cap procurement problem with conditionally i.i.d. sellers and a risk-neutral expected-profit objective.}
It does not model bundles, inventory, repeated interaction, endogenous seller strategies, or dynamic bargaining. 
.
Acceptance curves must be learned from offer--response data rather than listing prices, and seller heterogeneity, correlation, and legal commitment differences are treated only through reduced-form extensions. 
We evaluate the economic controller, not LLM persuasion quality.

\textit{The single negotiation window is essential.}
If failed negotiations could be retried immediately at negligible cost, sequential launch would weakly dominate parallel launch. 
Our results therefore apply only when all negotiations must be initiated before terminal responses arrive. 
Multi-window settings require dynamic batch-size decisions.

\textit{The common-cap abstraction treats all prices below \(b\) as operationally equivalent.}
Real procurement may jointly optimize price, delivery, warranty, or other contract attributes, and multiround bargaining may settle below the cap. 
Extending the decision from one scalar to multiple attributes would require multidimensional acceptance models.

Finally, the factorial grid is designed to cover representative market regimes rather than reflect their empirical frequency. 
Its normalization enables relative comparisons but does not calibrate \(\kappa\) or \(\delta\) to currency units. 
Practical deployment therefore requires parameter estimation, periodic recalibration, and robust fallbacks under distribution shift.

\subsection{Responsible Use}
Parallel agents can amplify spam, bargaining power asymmetry, deception, accidental commitments, and seller-contact externalities. 
Deployments should therefore rate-limit seller contact, disclose automation where required, distinguish tentative from binding actions, enforce price caps outside the language model, maintain auditable logs, and provide human oversight for high-stakes cases. 
Acceptance estimation should minimize personal data, while seller-level rate limits and cooldowns should constrain \(N\) before optimization, and \(\kappa\) should reflect seller-attention costs as well as computation. 
Applicable contract, consumer, privacy, and anti-\linebreak[4]discrimination law always take precedence over economic objectives.


\subsection{Conclusion.}
Agentic replication changes the economics of negotiation. Its benefit is not simply a higher probability of agreement: Parallelism permits lower per-conversation concession, but only until geometric fulfillment gains are overtaken by thread and commitment costs at a fixed target. 
\CANO{} jointly re-optimizes the target and turns that tradeoff into an auditable policy, and the dispersion sign reversal identifies when a varied market is an opportunity rather than a liability. 
These results support a design principle for agentic commerce: delegate language, but centrally optimize and enforce economic authority.


\bibliographystyle{ACM-Reference-Format}
\bibliography{references}

\clearpage

\appendix
\section{Derivations and Complete Proofs}
\label{app:proofs}

\subsection{Expected Objective}
\label{app:derivation}

For \(m\geq1\), let \(K=\sum_{i=1}^m\ind\{R_i\leq b\}\). Conditional on the
product state, \(K\sim\operatorname{Binomial}(m,q)\), where \(q=F(b)\).
The realized payoff is
\[
U(K)=
\begin{cases}
-L-\kappa m, & K=0,\\
V-b-\kappa m-\delta(K-1), & K\geq1.
\end{cases}
\]
The excess-commitment identity follows pointwise from
\((K-1)^+=K-\ind\{K\geq1\}\):
\[
\E[(K-1)^+]=\E[K]-\Prob(K\geq1)=mq-S_m(q).
\]
Therefore
\begin{align*}
\E[U(K)]
&=(V-b)S_m(q)-L[1-S_m(q)]-\kappa m\\
&\quad-\delta[mq-S_m(q)]\\
&=[V+L+\delta-b]S_m(q)-L-\kappa m-\delta mq.
\end{align*}
Substituting \(b=C(q)\) yields~\eqref{eq:profit-q}.

The convention \(m=0\) represents immediate cancellation after the customer
order has been accepted, so \(\Pi(0)=-L\). For \(m\geq1,q=0\), procurement
fails surely and \(\Pi(m,0)=-L-\kappa m\), weakly worse than cancellation. For
\(q=1\) when \(C(1)<\infty\), every launched thread accepts and
\(\Pi(m,1)=V-C(1)-\kappa m-\delta(m-1)\); hence \(m=1\) dominates larger
concurrency at this boundary. For unbounded support, no finite cap attains
\(q=1\); optimization is over \(q\in[0,1)\), with the upper boundary interpreted
by a limit.
\hfill$\blacksquare$

\subsection{Proof of Proposition~\ref{prop:threshold}}
\label{app:proof-threshold}

Since
\[
S_{m+1}(q)-S_m(q)=q(1-q)^m
\]
and the expected acceptance count rises by \(q\), subtracting
\(\Pi(m,q)\) from \(\Pi(m+1,q)\) gives~\eqref{eq:marginal}. Write
\(G(q)=V+L+\delta-C(q)\). If \(G(q)>0\), then
\[
\Delta_{m+1}(q)-\Delta_m(q)
=-G(q)q^2(1-q)^m<0.
\]
If \(G(q)\leq0\), the first term of~\eqref{eq:marginal} is nonpositive, so
\(\Delta_m(q)\leq-\kappa-\delta q\leq0\). Thus, over positive integers, profit
rises while \(\Delta_m(q)>0\), is flat when \(\Delta_m(q)=0\), and falls after
the marginal becomes negative. Resolving a zero marginal toward the smaller
thread count gives the smallest \(m\) with \(\Delta_m(q)\leq0\); when no such
\(m<N\) exists, the solution is \(N\).

For comparative statics, when \(G(q)>0\),
\[
\frac{\partial\Delta_m}{\partial V}
=\frac{\partial\Delta_m}{\partial L}
=q(1-q)^m>0,\qquad
\frac{\partial\Delta_m}{\partial\kappa}=-1,
\]
and
\[
\frac{\partial\Delta_m}{\partial\delta}
=q(1-q)^m-q=-q[1-(1-q)^m]\leq0.
\]
Moving every marginal in one direction moves its crossing threshold weakly in
the same direction. \(N\) only truncates the resulting threshold. 
\hfill$\blacksquare$

\subsection{Proof of Theorem~\ref{thm:substitution}}
\label{app:proof-substitution}

For compactness, set
\[
B(q)=V+L+\delta-C(q).
\]
The first two derivatives of the fixed-\(m\) objective are
\begin{align*}
\Pi_m'(q)
&=-C'(q)S_m(q)+B(q)S_m'(q)-\delta m,\\
\Pi_m''(q)
&=-C''(q)S_m(q)-2C'(q)S_m'(q)+B(q)S_m''(q).
\end{align*}
On the interval in Assumption~\ref{ass:regular},
\(S_m'(q)=m(1-q)^{m-1}>0\), \(S_m''(q)\leq0\), \(B(q)\geq0\),
\(C'(q)>0\), and \(C''(q)\geq0\). Hence
\(\Pi_m''(q)<0\), proving strict concavity and uniqueness.

At an interior optimum, \(\Pi_m'(q)=0\). Dividing by
\(S_m'(q)=m(1-q)^{m-1}\) gives
\[
V+L+\delta-C(q)
=C'(q)A_m(q)+\delta(1-q)^{-(m-1)}.
\]
Rearrangement yields~\eqref{eq:foc}.

It remains to compare roots. Let
\[
H_m(q)=C(q)+C'(q)A_m(q)
+\delta[(1-q)^{-(m-1)}-1].
\]
For fixed \(q\in(0,1)\), write \(r=1-q\). Then
\[
A_m(q)=\frac{q}{m}\sum_{j=0}^{m-1}r^{-j}.
\]
The sequence \(1,r^{-1},r^{-2},\ldots\) is strictly increasing, so appending
its next term raises its running average:
\(A_{m+1}(q)>A_m(q)\). The commitment term is also nondecreasing in \(m\).
Thus \(H_{m+1}(q)\geq H_m(q)\).

For fixed \(m\), \(A_m(q)\) is strictly increasing because both \(q\) and every
\(r^{-j}\) are nondecreasing and the \(j=0\) term makes the product strict.
Together with \(C'>0,C''\geq0\), this makes \(H_m(q)\) strictly increasing.
Since both optima solve \(H_m(q)=V+L\),
\[
q_{m+1}^\star\leq q_m^\star.
\]
Finally \(C\) is increasing, so the same inequality holds for the caps.
\hfill$\blacksquare$

\subsection{Proof of Theorem~\ref{thm:dispersion}}
\label{app:proof-dispersion}

For fixed \(m\), only the quantile curve in~\eqref{eq:profit-q} depends on
\(\sigma\). At a unique optimum, the envelope theorem gives
\[
\frac{d\Pi_m^\star(\sigma)}{d\sigma}
=-\frac{\partial C_\sigma(q_m^\star)}{\partial\sigma}
S_m(q_m^\star)
=-z(q_m^\star)S_m(q_m^\star).
\]
Because \(S_m(q)>0\) for \(q>0\) and symmetry implies
\(z(q)<0\) below \(1/2\), \(z(1/2)=0\), and \(z(q)>0\) above \(1/2\), the
three sign cases follow. If a positive joint optimizer over \(m\) is unique in
a neighborhood of \(\sigma\) and its fixed-\(m\) solution remains unique and
interior, the same envelope derivative applies to the active \(m\). At an
integer switch, the upper envelope of the fixed-\(m\) value functions is
continuous but need not be differentiable. \hfill$\blacksquare$

\section{\CANO{} Algorithm and Extensions}
\label{app:algorithm}

\subsection{Planner Output and Failure Behavior}
\label{app:output}
The planner receives a market instance containing
\((N,V,L,\kappa,\delta)\) together with the estimated quantile function
\(C(q)\). It returns either a cancellation decision or an optimal policy
tuple
\[
(m^\star,q^\star,b^\star),
\]
while fulfillment and commitment metrics are evaluated separately by a
deterministic evaluator.

The implementation first performs a bounded global search over
\(q\in[10^{-6},1-10^{-6}]\), followed by golden-section refinement around
the best candidate. For finite-support distributions, the boundary point
\(q=1\) is additionally checked. The resulting policy is finally compared
with immediate cancellation, and cancellation is selected whenever every
positive-concurrency policy yields a lower expected value than \(-L\).

On a reference uniform instance, the implementation matches an exhaustive
100,001-point grid search to within \(10^{-8}\). This serves as a regression
check rather than a universal optimality guarantee. Theoretical global
optimality remains guaranteed only under
Assumption~\ref{ass:regular}. Implementation details are provided in the following.

\subsection{Algorithm}

For each \(m\in\{1,\ldots,N\}\), \CANO{} performs:
\begin{enumerate}
  \item Construct \(q_g=\underline q+g(\overline q-\underline q)/(G-1)\) for
  \(g=0,\ldots,G-1\).
  \item Evaluate \(\Pi(m,q_g)\) and retain the best cell and both neighbors.
  \item Refine that bracket by golden-section maximization to tolerance
  \(\varepsilon\), also checking global endpoints.
  \item Convert the best \(q_m\) to \(b_m=C(q_m)\) and record its value.
\end{enumerate}
It returns the best recorded positive-\(m\) policy unless cancellation
\(-L\) is larger. Under Assumption~\ref{ass:regular}, Step 1 may be replaced by
boundary derivative checks and bisection of~\eqref{eq:foc}. Strict concavity
proves global correctness, while bisection needs
\(O(\log(1/\varepsilon))\) evaluations for each \(m\). The global-scan
implementation is intentionally more conservative for empirical curves.

\subsection{Correlated Sellers}
\label{app:correlation}

The i.i.d. theory does not apply when seller reservation prices share shocks.
For the stress test, let latent acceptance variables follow an exchangeable
Gaussian copula:
\[
Z_i=\sqrt{\rho}Z_0+\sqrt{1-\rho}\epsilon_i,
\]
where all normal factors are independent and a seller accepts when
\(Z_i\leq\Phi^{-1}(q)\). Conditional on \(Z_0=z\), acceptance probability is
\[
p_q(z)=\Phi\!\left(
\frac{\Phi^{-1}(q)-\sqrt{\rho}z}{\sqrt{1-\rho}}
\right).
\]
Consequently,
\[
S_m^\rho(q)=1-\E_{Z_0}[(1-p_q(Z_0))^m].
\]
Marginal acceptance remains \(q\), so \(\E[K]=mq\) and the objective retains
the form
\[
\Pi_\rho(m,q)=[V+L+\delta-C(q)]S_m^\rho(q)
-L-\kappa m-\delta mq.
\]
We evaluate the one-dimensional expectation with 32-node Gauss--Hermite
quadrature, enumerate \(m\), and maximize over a 513-point \(q\)-grid. Positive
dependence lowers the diversification benefit of concurrency, explaining the
shift toward fewer, more accommodating threads.

\begin{figure}[!t]
  \centering
  \includegraphics[width=0.5\textwidth]{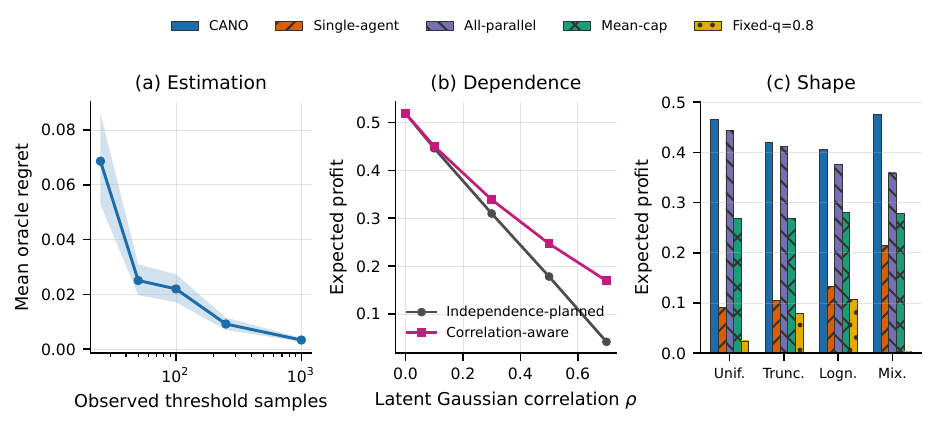}
  \caption{Stress tests. (a) Mean regret and 95\% bootstrap interval over 100
  finite-data replicates. (b) Correlation-aware planning partially recovers
  profit lost under shared seller shocks. (c) Policy profit under uniform,
  truncated-normal, lognormal, and two-component reservation-price
  distributions.}
  \Description{Three panels. Mean oracle regret falls from about 0.069 with 25
  observed threshold samples to 0.003 with 1,000. As the tested latent Gaussian
  correlation rises from 0 to 0.7, expected profit falls; at every positive
  tested correlation, the correlation-aware line is above the
  independence-planned line. Hatched grouped bars compare five policies under
  four mean-one, standard-deviation-about-0.30 distribution shapes.}
  \label{fig:robustness}
\end{figure}

\section{Additional Experimental Detail}
\label{app:experiments}

\subsection{Full Protocol}

All results are generated by one Python script using NumPy and Matplotlib. The
main seed is 20,260,803. Analytic rows, figures, and LaTeX result macros are
written from the same run. The implementation contains separate modules for
distributions, the profit model, optimization, correlation, and simulation.
No proprietary data, model API, or human-subject data are used.

\begin{table}[t]
  \caption{Factorial design for the \(\GridScenarioCount{}\) analytic markets.}
  \label{tab:grid}
  \centering
  \small
  \begin{tabular}{@{}ll@{}}
    \toprule
    Parameter & Values \\
    \midrule
    \(N\) & \(1,2,5,10,20,50\) \\
    \(\sigma/\mu\) & \(.05,.15,.30,.45\) \\
    \(V-\mu\) & \(.10,.30,.50\) \\
    \(L\) & \(0,.10,.30\) \\
    \(\kappa\) & \(.001,.01,.03\) \\
    \(\delta\) & \(0,.02,.10\) \\
    \bottomrule
  \end{tabular}
\end{table}

The primary metrics are expected profit, fulfillment probability, cap,
concurrency, expected excess commitments \(mq-S_m(q)\), duplicate probability
\[
\Prob(K\geq2)=1-(1-q)^m-mq(1-q)^{m-1},
\]
and absolute regret. Fixed-\(q=.7\) and \(.9\) sensitivities, as well as
one-factor sweeps in \(V,L,\kappa,\delta\), are included in the raw output.

\subsection{Monte Carlo Validation}

For \(N=20,\mu=1,\sigma=.35,V=1.4,L=.2,\kappa=.01,\delta=.05\), \CANO{}
selects \(m=14,q=.15097,b=.57682\). Analytic expected profit is \(0.518956\).
Across one million independent orders, simulated mean profit is \(0.519354\)
with standard error \(0.000293\); absolute error is \(0.000397\). Simulated
fulfillment and duplicate probabilities are \(0.89924\) and \(0.64661\).

\section{Robustness Results}
\label{app:robustness}

\subsection{Distribution Shape}

Table~\ref{tab:shape} reports the policies underlying
Figure~\ref{fig:robustness}(c). All four distributions have mean one and
standard deviation approximately \(0.30\). The truncated normal is symmetric
at three underlying standard deviations. The two-component market contains a
low-price segment and a smaller high-price segment; its quantile is represented
by 100,000 fixed-seed draws and captures unobserved seller-type heterogeneity.
Because some shapes violate the convex-quantile condition globally, the bounded
grid scan followed by local refinement is used throughout.

\begin{table}[t]
  \caption{Distribution robustness: \(m\), \(q\), and expected profit.}
  \label{tab:shape}
  \centering
  \scriptsize
  \begin{tabular}{@{}lrrrr@{}}
    \toprule
     & \multicolumn{2}{c}{Uniform} & \multicolumn{2}{c}{Trunc.\ normal}\\
    Policy & \(m,q\) & Profit & \(m,q\) & Profit\\
    \midrule
    \CANO{} & \(13,.161\) & .465 & \(15,.134\) & .421\\
    Single & \(1,.539\) & .092 & \(1,.627\) & .105\\
    All & \(20,.117\) & .444 & \(20,.106\) & .412\\
    Mean-cap & \(4,.500\) & .269 & \(4,.500\) & .269\\
    Fixed-.8 & \(2,.800\) & .025 & \(2,.800\) & .079\\
    \bottomrule
  \end{tabular}

  \medskip
  \begin{tabular}{@{}lrrrr@{}}
    \toprule
     & \multicolumn{2}{c}{Lognormal} & \multicolumn{2}{c}{Mixture}\\
    Policy & \(m,q\) & Profit & \(m,q\) & Profit\\
    \midrule
    \CANO{} & \(11,.187\) & .406 & \(5,.429\) & .477\\
    Single & \(1,.641\) & .133 & \(1,.540\) & .215\\
    All & \(20,.115\) & .377 & \(20,.133\) & .360\\
    Mean-cap & \(3,.558\) & .280 & \(3,.550\) & .278\\
    Fixed-.8 & \(2,.800\) & .107 & \(2,.800\) & .003\\
    \bottomrule
  \end{tabular}
\end{table}

\subsection{Parameter Comparative Statics}

The numerical one-factor sweeps agree with Proposition~\ref{prop:threshold}.
In the reference uniform market, raising \(\kappa\) from .001 to .06 reduces
\(m^\star\) from 20 to 4 and raises \(q^\star\) from .117 to .325. Raising
\(\delta\) from 0 to .20 reduces \(m^\star\) from 15 to 9 and expected excess
commitments from 1.659 to .637. Raising \(L\) from 0 to .60 increases
concurrency from 12 to 14 and fulfillment from .869 to .929. Because \(L\)
also appears in the failure payoff, reported profit need not increase with
\(L\); the policy nevertheless becomes more reliable as predicted.

\subsection{Scope of the Agentic Evaluation}

The experiment targets the economic control layer shown in
Figure~\ref{fig:architecture}. It samples seller acceptance and commitment
outcomes but does not call an LLM, measure rhetoric, or compare prompts.
Accordingly, no claim in the paper attributes profit improvements to language
generation. A future controlled LLM evaluation should pair identical market
seeds, seller messages, and model snapshots; distinguish attempted from
executed cap violations; and report token use, latency, and prevented duplicate
commitments. The current results establish the policy that such a language
layer should be constrained to execute.

\end{document}